\documentclass [conference]{IEEEtran}
\usepackage[utf8]{inputenc}
\usepackage{graphicx}
\usepackage{geometry}
\usepackage{float}
\usepackage{mathtools}
\usepackage{subfig}
\usepackage[table,xcdraw]{xcolor}
\usepackage{lettrine}
\usepackage{nopageno}
\usepackage{algorithm,algorithmic}
\usepackage{url}
\usepackage{cite}
\usepackage{flushend}
\usepackage{longtable}
\usepackage{multirow}
\usepackage{tikz}
\usepackage{siunitx}
\title{An Optimal Multi-UAV Deployment Model \\ for UAV-assisted Smart Farming} 

\author{\IEEEauthorblockN{Shavbo Salehi}
\IEEEauthorblockA{\textit{Electrical and Computer Engineering } \\
\textit{Department, Urmia University}\\
Urmia, Iran \\
shavbo.salehi@urmia.ac.ir}
\and
\IEEEauthorblockN{Jahan Hassan}
\IEEEauthorblockA{\textit{School of Engineering and Technology} \\
\textit{The Central Queensland University}\\
Sydney, Australia \\
j.hassan@cqu.edu.au}
\and
\IEEEauthorblockN{Ayub Bokani}
\IEEEauthorblockA{\textit{School of Engineering and Technology} \\
\textit{The Central Queensland University}\\
Sydney, Australia \\
a.bokani@cqu.edu.au}
}

\begin{document}
\maketitle

\begin{abstract}

Next-generation wireless networks will deploy UAVs dynamically as aerial base stations (UAV-BSs) to boost the wireless network coverage in the out of reach areas. To provide an efficient service in stochastic environments, the optimal number of UAV-BSs, their locations, and trajectories must be specified appropriately for different scenarios. Such deployment requires an intelligent decision-making mechanism that can deal with various variables at different times. This paper proposes a multi UAV-BS deployment model for smart farming, formulated as a Multi-Criteria Decision Making (MCDM) method to find the optimal number of UAV-BSs to monitor animals' behavior. This model considers the effect of UAV-BSs' signal interference and path loss changes caused by users' mobility to maximize the system's efficiency. To avoid collision among UAV-BSs, we split the considered area into several clusters, each covered by a UAV-BS. Our simulation results suggest up to 11x higher deployment efficiency than the benchmark clustering algorithm.

\end{abstract}


\section{Introduction}
\label{section:intro}

The Fifth Generation (5G) mobile networks aims to support many use cases ranging from autonomous vehicles, e-health, industry 4.0, and smart cities, leading to a noticeable increase in network demands. This increase has seen emerging cutting-edge services, such as Unmanned Aerial Vehicles (UAVs), attracting research and industry interest. 
Next-generation networks utilize UAVs as aerial base stations (UAV-BSs) to provide services through dynamic wireless coverage in low bandwidth areas. These networks are to support the ever-increasing Internet of Things (IoT) applications~\cite{SALEHI2021108415,our_ccnc_traj}. However, an efficient service delivery mechanism requires proper positioning of UAV-BSs to establish reliable communication channels with ground nodes. On the other hand, to reduce the overall deployment costs and have a minimal number of UAV-BSs, they require an optimal trajectory design to cover all users efficiently. Such trajectories will depend on multiple effective parameters, varying for different scenarios. 

In this paper, we consider a scenario of several UAV-BSs acting as relay nodes to provide fast and reliable connectivity between mobile ground nodes (farm animals tagged with IoT devices) and a terrestrial base station in a \textit{smart farm}. This study aims to provide the maximum coverage by the minimal number of \textit{available UAVs} in the network while ensuring collision-free and seamless operations of the UAVs in conventional air traffic. For this purpose, we design a  Multi-Criteria Decision Making (MCDM) mechanism for UAV-BSs deployment that serves mobile ground nodes with different service demands while avoiding UAV collisions and minimizing the total costs. We consider ground IoT nodes with various delay-tolerant service demands that require different bandwidth ranges. Additionally, to serve varying numbers of mobile ground nodes, we propose utilizing several UAV-BSs instead of a single unit. However, this introduces additional challenges which require the following points to be addressed:

\begin{itemize}
    \item Interference management
    \item Determining the optimal number of UAV-BSs
    \item Designing a cooperative UAV trajectory
\end{itemize}
We define the most suitable minimal number of UAV-BSs by (i) the system interference level due to using two types of BSs (terrestrial and aerial), (ii) mobile ground nodes' requirements, (iii) the distance between the nodes and UAVs, and (iv) the interference effects of utilizing more number of UAVs in the region. At the same time, our MCDM method aims to achieve a collision-free UAV-BSs deployment. We also investigate different practical scenarios for deriving the assumptions in our studied system.

Since UAVs rely on their on-board batteries for power, their flight duration and coverage area are limited. Therefore, we aim to find an optimal number of UAVs to provide the highest possible coverage. However, to prevent UAVs collisions and interference, our method allows each UAV-BS to serve only one cluster of animals at a time. We propose using the MCDM method to meet the bandwidth and delay requirements of the maximum number of ground nodes while minimizing the number of UAVs in the system by utilizing a dynamic k-Means clustering method. Simulation results demonstrate that our proposed MCDM method can significantly reduce the required number of UAV-BSs compared to the benchmark model for most farm sizes and similar numbers of mobile nodes.



\section{Related Work}
\label{related-works}


Researchers investigated several key issues, including adjusting the UAVs altitudes \cite{SALEHI2021108415}, energy consumption \cite{salehi2019aetd,OurISPN, jahan_recharging}, and trajectory design \cite{tran2020coarse}, to facilitate the deployment of networked UAVs. A multi-UAV trajectory design is proposed in \cite{chang2020machine} by a combination of reinforcement learning and a deep learning algorithm to tackle resource allocation problems. In \cite{zhao2020multi} trajectory design and power allocation challenges of UAVs are addressed by a deep reinforcement learning model in multi-UAV networks. In \cite{zhang2020three} a Deep Q-Network algorithm formulated the UAVs' coverage constraint issues due to three-dimensional (3D) mobility.

Although utilizing multiple UAVs in a large region sounds promising, it increases the overall service delivery costs. Adapting the number of UAVs to the number of their targeted mobile ground nodes can be a viable solution to reduce deployment costs. In \cite{reina2018multi} researchers showed how UAV-assisted networks managed by a multi-layout multi-sub-population genetic algorithm could outperform the heuristic models when providing services to the various number of ground nodes in a large area. Additionally, researchers proposed several methods for serving an area with a fair number of UAVs, e.g., in \cite{8485481} Nash Equilibrium is proposed for finding an energy-efficient trajectory and increasing the coverage area for multi-UAV. In \cite{li2019reactive} livestock is tracked and monitored by the minimum number of UAVs with a density-based clustering algorithm.


Multi UAVs were employed to provide services to industrial IoT devices, monitoring sensors in farms, and livestock tracking. While UAVs have been used widely in smart farming, utilizers have faced several challenges, including limited energy resources and collisions among UAVs due to the size of farms. Since the farms are broad, a group of UAVs must monitor them. In this paper, we propose utilizing the MCDM method to determine the minimal number of UAVs required for monitoring mobile ground nodes for reducing the cost of monitoring and the probability of collision. The considered ground nodes' mobility follows the Random Waypoint mobility model, and they are distributed in three different-sized farms. The mobile users lead to dynamic clustering in the farm, then intending to decrease the probability of collisions among UAV-BSs, the farm is clustered to disjoint areas. Additionally, due to the limited batteries installed on UAVs and GPS devices, the MCDM method is optimized to have the lowest energy consumption alongside the highest area coverage. 

\section{Proposed Method}
\label{proposed-method}



We propose an MCDM model, which allows UAV-BSs to provide various services to mobile, dynamic, and randomly distributed ground IoT devices. In this model, we first classify ground nodes into multiple clusters based on their current locations by using the $k$-means algorithm. Then, a UAV-BS is assigned to each cluster. However, finding a safe separation between UAVs is essential to avoid signal interference. This criterion leads to considering another variable in the clustering model. Because we consider dynamic mobile ground nodes in real-world scenarios, we need to re-cluster and re-do the UAV allocations periodically. Therefore, an optimal number of UAV-BSs must be assigned based on ground mobile nodes' density and distribution pattern. 

\begin{figure}[hbt!]
   \centering
  \includegraphics[width=\linewidth]{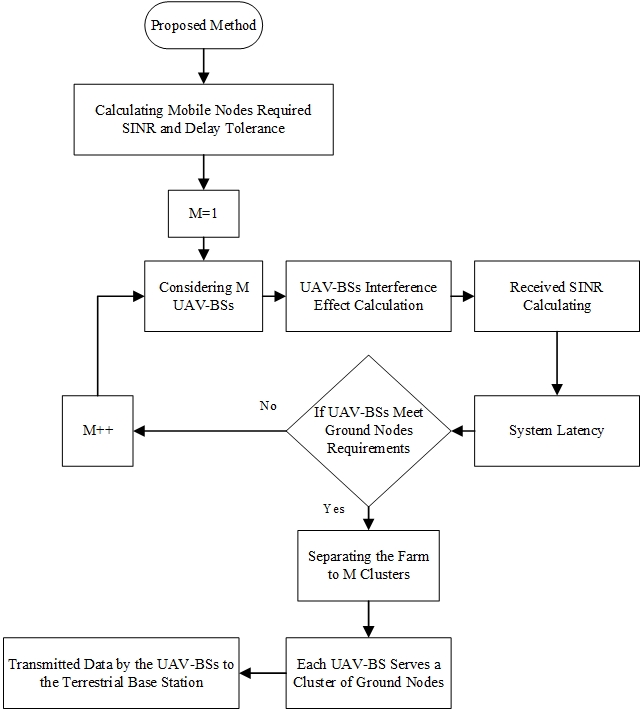}
   \caption{The Multi-Criteria Decision Making approach flowchart}
     \label{fig:flowchart}
 \end{figure}

This section briefly explains the considered scenario, obtaining the optimum number of UAVs, and defining the distance to avoid interference and collisions in our novel clustering model. Fig. \ref{fig:flowchart} provides an overall perspective of the method.
 
\subsection{Scenario}
\label{subsection:scenario}
 
\begin{figure}[hbt!]
  \centering
9/6  \includegraphics[width=0.65\linewidth]{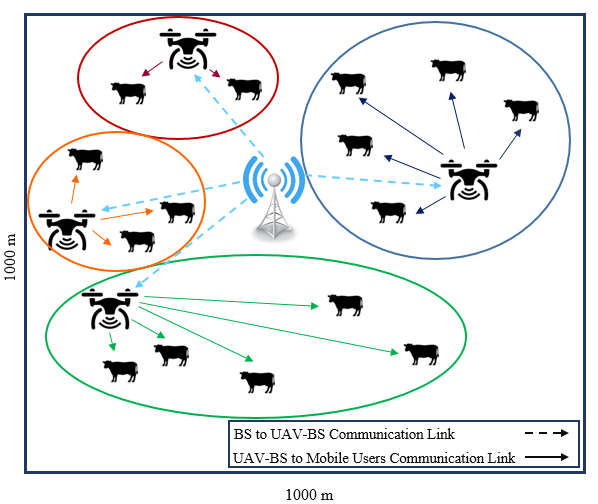}
  \caption{Mobile nodes clusters and allocated UAV-BSs}\label{fig:sce2}
\end{figure}

We consider various smart farming scenarios in which several farm animals equipped with GPS and health monitoring sensors (mobile ground nodes or users) are distributed randomly in the farmland area. The nodes require distinct services, then UAV-BSs are deployed to collect nodes' data and send it to a terrestrial BS located on the farm. Fig. \ref{fig:sce2} illustrates mobile ground nodes randomly distributed in the farm that are clustered based on their current locations. The animals' mobility follows the Random Waypoint model. Due to mobility, the clusters' sizes change continuously. 

We attempt to indicate the effect of the farmland area and the number of animals on meeting users’ bandwidth and delay requirements by changing the number of UAV-BSs serving the area. In practical scenarios, based on the number of ground nodes distribution, density, and their required services collected by the terrestrial BS, various numbers of UAV-BSs are utilized to cover the area. It must be noted that UAVs with megahertz waves (MHz) communication capability serve ground nodes. However, nodes receive a Signal-to-interference-plus-noise (SINR) ratio that decreases due to increasing the distance between the mobile nodes and both types of BSs due to path loss, interference, and noise. 

\subsection{Optimization Targets and Constraints}

UAVs' limited flight duration hinders their capability to meet mobile users' requirements when users are distributed in larger areas, e.g., farmland, since a considerable percentage of UAVs energy is consumed for flying \cite{SALEHI2021108415}. Then, utilizing Multi UAVs is suggested for meeting users' requirements. Nevertheless, this leads to an increased probability of UAVs crashing and collisions. Furthermore, serving nodes by several UAVs is another challenge due to the interference of multiple UAV-BSs data transmission.

Therefore, we endeavor to cover various farmlands with an optimal number of UAVs for meeting users' requirements. To avoid UAVs collisions and improve received service quality, we separate the area based on the number of UAVs for avoiding collisions. Additionally, each UAV-BS serves an area where several ground nodes are required to be served. Intending to meet the services' requirements, the MCDM method is proposed for covering the maximum number of ground nodes with a minimal number of UAVs by utilizing a dynamic K-Mean clustering method. 






\subsection{Finding the Optimum Number of UAVs}
\label{subsection:MCDN}

To find an optimal number of UAVs that meet ground nodes' requirements, we consider the effects of distinct criteria by Multi-criteria Decision Making (MCDM) method:

\begin{itemize}
  \item The interference level is related to the distance between UAV-BSs.  
  \item Effect of path loss on the signal strength related to the distance between mobile ground nodes and UAV-BSs.
\end{itemize}


The proposed MCDM method specifies the feasible solution according to the above-established criteria, including SINR, path loss, and interference that affect the quality of the mobile communication for meeting users' bandwidth and delay requirements.

\subsubsection{Interference Management of Multi UAV-BS system}
\label{subsubsection:Interferencemanag}

Similar to the heterogeneous networks (HetNets), by increasing the number of UAV-BSs, the probability of interference and collision increases. In this section, we endeavor to find the lowest number of UAVs to serve the maximum number of nodes and decrease the UAVs' energy consumption, the probability of collisions and increase the quality of services. First, the received SINR is calculated in UAV-BSs, which are connected directly to the terrestrial BS on the sub-carrier $n$ as:

\begin{equation}
SINR_{M_{BS},n} = \frac{P_{d}^{M_{BS}}G_{M_{BS},n}\omega d^{-\eta}}{NI}
\label{eqn:eq1}
\end{equation}
where $P_{d}^{M_{BS}}$ denotes the received signal power on the UAV-BS at a distance $d$ from the BS in the considered cell, $G_{M_{BS}}$ is the channel gain, and NI is calculated by the equation below:

\begin{multline}
NI = \\
\sum_{i=1,i\not=M_{BS}}^{M_{BS}'} P_{d}^{M_{BS}'}G_{M_{BS}',n} + \sum_{j=1}^{U_{BS}'} P_{d}^{U_{BS}}G_{U_{BS},n} + P_{noise}^{\Delta f}
\label{eqn:eq2}
\end{multline}
where $P_{d}^{M_{BS}'}$ is the received signal power from the adjacent BS in the first tier, $G_{M_{BS}'}$ denotes the gain of other channels, and $P_{noise}^{\Delta f}$ is the white noise power. Moreover, the received SINR is calculated in mobile users which are connected to the UAV-BS directly on the subcarrier $n$ by the equation below:

\begin{equation}
SINR_{U_{BS},n} = \frac{P_{d}^{U_{BS}}C_{Loss}}{NI_{II}}
\label{eqn:eq3}
\end{equation}

\begin{multline}
NI_{II} = \\
\sum_{i=1,i\not=M_{BS}}^{M_{BS}'} P_{d}^{M_{BS}'}G_{M_{BS}',n} + \sum_{j=1}^{U_{BS}} P_{d}^{U_{BS}}G_{U_{BS},n} + P_{noise}^{\Delta f}
\label{eqn:eq4}
\end{multline}

Ground nodes equipped with various sensors for data gathering may require various bandwidth and delay to send data to the UAVs. The most suitable number of UAVs are assinged for serving nodes to meet service requirements. Furthermore, the mobile ground nodes requiring resource blocks $RB$ to evaluate the required bandwidth ($BW$) are calculated by the equation below:


\begin{equation}
BW = \sum_{n=1}^{N_{Tot}} RB 
\end{equation}
where $N_{Tot}$ shows the number of active nodes requiring $RB$, which meet by the UAV-BSs, intending to consider the service reliability and delay tolerance, the ground nodes' requested services are delivered at a particular time. For this reason, the total service latency ($L$) is calculated by considering the service delay at every single node ($l$) as $L = \sum_{n=1}^{N_{Tot}} l$.

The required UAV-BSs for a particular area are selected based on multiple parameters, including the number of mobile users, service delivery latency, nodes' bandwidth requirements, and the size of the area. Regarding service delivery latency, we consider nodes requiring services that must be met by UAV-BSs before the deadline (or latency). In table \ref{tab:comperinf}, the effect of using multi UAV-BSs on meeting service requirements is listed.

Overall, by increasing the number of UAV-BSs, the interference, the probability of UAV-BSs collision, and deployment costs increase. Therefore, deploying an optimal number of UAV-BSs in the farmland increases system efficiency significantly. 

\subsubsection{Path Loss Management of Multi-UAVs}
\label{subsubsection:pathloss}

Despite the various benefits of radio frequencies in mobile communication, they suffer from quality degradation caused by path loss. Consequently, radio propagation models to predict the cellular transmission path loss is proposed to deal with this problem. In this paper, we use the Okumura–Hata model to indicate the path loss, which is calculated by the equation below: 

\begin{multline}
 PL_{(dB)} = L_{U_db} - 4.78 (\log_{10} f_{MHz})^2 \\ 
 + 18.33 (\log_{10} f_{MHz})^2 - 40.94
 \label{eqn:eq5}
\end{multline}
where $PL$ denotes the path loss, $L_U$ is the average path loss in small cities, and $f$ is transmission frequency. This paper considers the profound effects of path loss on the quality of received services.

It should be noted that in eq(\ref{eqn:eq1}), $P_{d}^{M_{BS}}$ provides information regarding the received signal in UAV-BS from the terrestrial BS (between two types of BSs). In contrast, eq(\ref{eqn:eq5}) apprise the path loss effects on the received signal in ground nodes (between UAV-BS and users).

\subsection{K-Means}

\begin{figure}[hbt!]
  \centering
  \includegraphics[width=0.65\linewidth]{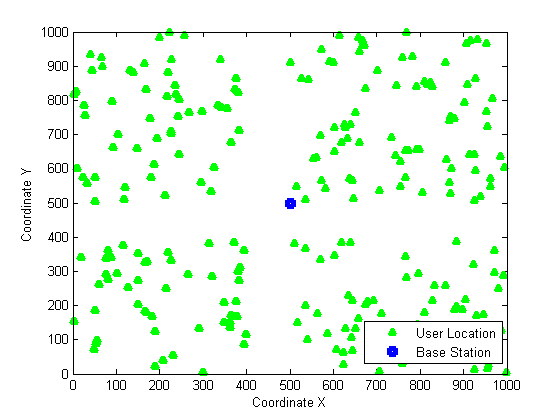}
  \caption{Mobile users distribution for 250 users in interest area}\label{fig:dis1}
\end{figure}

As mentioned in Section \ref{subsection:MCDN}, the number of UAV-BSs is selected by considering the ground nodes. Fig. \ref{fig:dis1} illustrates the various number of nodes distributed in the area randomly that move as per the Random Waypoint mobility model. We consider multiple scenarios, three different farmland sizes, and three various groups of mobile users to determine the best (optimal) number of UAV-BSs.

\begin{figure}[hbt!]
  \centering
  \includegraphics[width=0.65\linewidth]{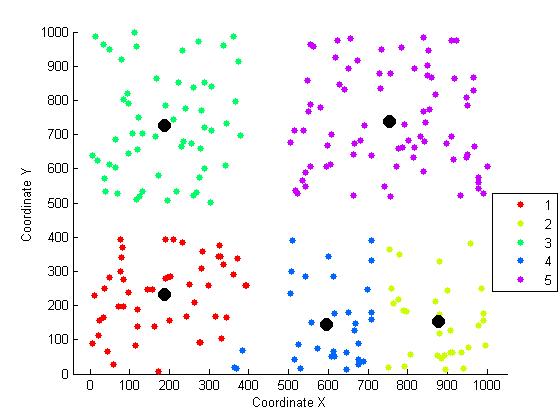}
  \caption{Mobile Users clustering by 5 UAV-BSs for 250 users}\label{fig:cl3}
\end{figure}

Fig. \ref{fig:cl3} depicts five UAV-BSs covering an area and serving nodes in assigned clusters. 
Moreover, Orchestrating all the UAVs requires the assumption that UAVs perceive the user's real-time position. In our clustering model, we prevent cluster overlapping from preventing UAV-BSs collision. In Fig. \ref{fig:cl3} clusters of nodes are shown in various colors, and the UAV-BSs are depicted by black circles.

\begin{table}[t]
\caption{Simulation Parameters}
    \centering
   \scriptsize
    \begin{tabular}{|l|l|}
\hline
MCDM based method Component & Value \\ \hline
$N_{Nodes}$ & 100, 200, and 500 \\
Cell Side & 1, 2, and 5 ($km$) \\ 
Frequency ($f$) & 1500 MHz \\ 
Terrestrial BS Channel Bandwidth & 20 MHZ \\
Transmission Power of Terrestrial BS & 45 dBm \\ 
Terrestrial BS Height & 100 m \\ 
Transmission Power of UAV & 20 dBm \\ 
UAV Flying Height & 100 m \\ 
Ground Nodes Speed (m/s) & [1,3] \\ 
Ground Nodes Height & 1.5 m \\
Ground Nodes Pause Interval & [0,1] \\
Ground Nodes Direction Interval & [-180,180] \\\hline
    \end{tabular}
    \label{tab:sim_param}
\end{table}

We calculate the number of clusters by considering the effect of user data transmission and signals leading to interference. The interference based on the Shannon theory, $Throughput=B\times log_2(1+\frac{S}{N+i})$, decreases the system throughput. Subsequently, fewer UAVs that meet the users' requirements must be used since interference is higher by increasing the number of UAVs. Furthermore, based on the path loss model, the signal strength decreases by increasing the distance between UAVs and users. Therefore, users' and UAV-BSs' distance and interference must be considered in cluster planning.

\section{Results}
\label{results}

We developed a custom simulator in MATLAB for studying the performance of our proposed MCDM method in various scenarios. All simulations are run 30 times, and the averages of these results are reported. Some predefined parameter values based on~\cite{zeng2019energy,li2019reactive} were used in the simulations as reported in Table~\ref{tab:sim_param}.

\subsection{Effects of the Number of UAVs' used on WCSS}

\begin{figure}[hbt!]
  \centering
  \includegraphics[width=0.8\linewidth]{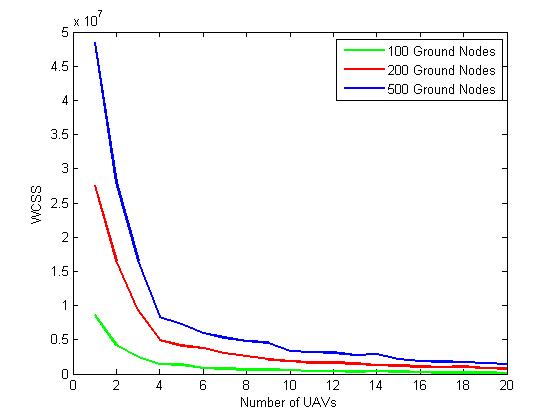}
  \caption{WCSS by various number of UAV-BSs for 100, 200, and 500 nodes.}
    \label{fig:wcss}
\end{figure}

In this paper, the number of clusters is aligned to the number of UAV-BSs in the various scenarios, intending to find the minimum number of UAV-BSs covering the area and meeting the nodes' requirements. To find the most suitable number of UAV-BSs, we calculate Within-Cluster Sums of Squares (WCSS), the sum of the squared distance between each node and the UAV-BSs as the centroid of clusters. The effects of various numbers of UAV-BSs on the distance between nodes and UAV-BSs are evaluated by WCSS. Fig.~\ref{fig:wcss} shows the WCSS in three distinct scenarios. Clearly, by increasing the number of nodes, the WCSS increases. It must be noted that we endeavor to provide a connection for nodes meeting their requirements. Consequently, the node's received signals are calculated by considering the path loss.

Since various radio frequencies have different coverage areas \cite{SALEHI2021108415}, we calculate the WCSS for evaluating the signal strength based on the requested services. 

\subsection{Effects of the number of UAVs on CQI}

\begin{figure*}[!htb]
\minipage{0.32\textwidth}
  \includegraphics[width=\linewidth]{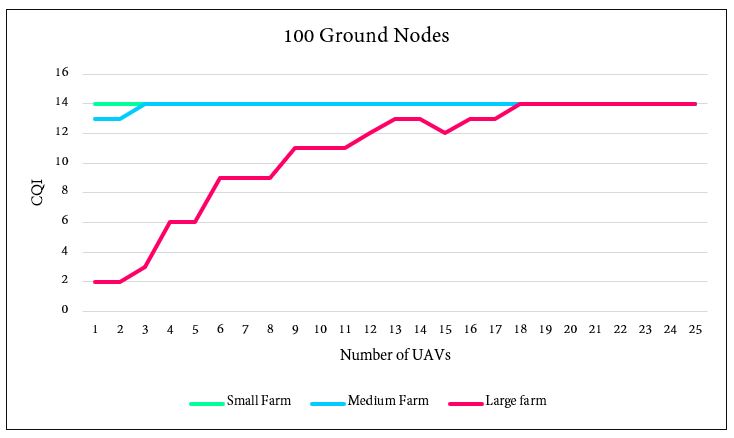}
  \caption{CQI by various number of UAV-BSs for 100 in three distinct sizes of farm}\label{fig:100}
\endminipage\hfill
\minipage{0.32\textwidth}
  \includegraphics[width=\linewidth]{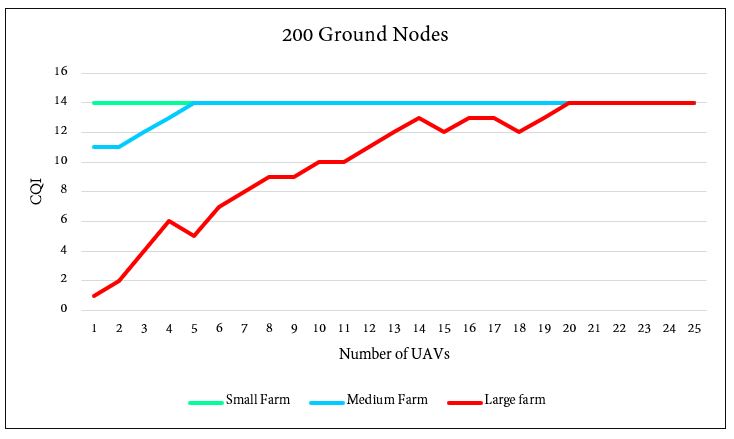}
  \caption{CQI by various number of UAV-BSs for 200 in three distinct sizes of farm}\label{fig:200}
\endminipage\hfill
\minipage{0.32\textwidth}%
  \includegraphics[width=\linewidth]{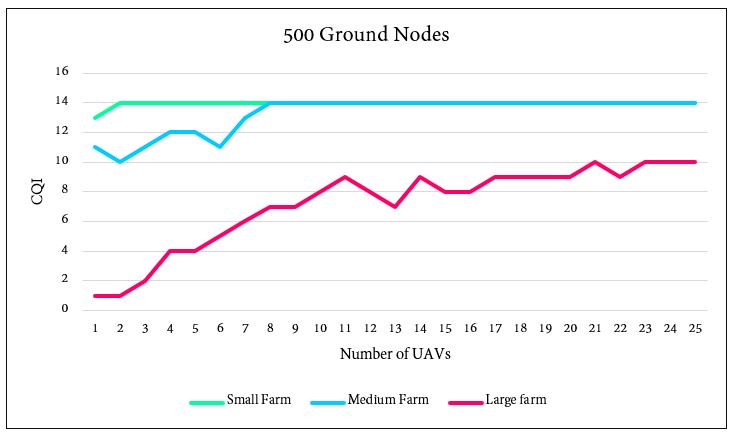}
  \caption{CQI by various number of UAV-BSs for 500 in three distinct sizes of farm}\label{fig:500}
\endminipage
\end{figure*}

As we discussed in Subsection ~\ref{subsubsection:Interferencemanag}, increasing the number of UAVs and ground nodes has an undeniable effect on the received SINR of users. Channel Quality Indication (CQI) is utilized for indicating the influence of the users' data transmission on channels. Overall, the CQI is classified into 15 different codes regarding the received SINR for Long-Term Evolution (LTE) telecommunications. Increasing the number of ground nodes has noticeable effects on the CQI, as shown in Figs.~\ref{fig:100} to~\ref{fig:500}. 

Figs.~\ref{fig:100}, ~\ref{fig:200}, and ~\ref{fig:500} illustrate that both the number of ground nodes and the size of the farms have profound effects on the CQI. Fig.~\ref{fig:100} provides information about the scenario with 100 nodes and one UAV-BS, which meets users' requirements. Moreover, by increasing the number of nodes to 200, the system requirements are still met by one UAV-BS (Fig. \ref{fig:200}). However, by increasing the number of nodes to 500, two UAV-BSs are required to serve the nodes with higher quality. Similarly, by extending the size of the farm, more UAV-BSs must be deployed to serve the nodes' requirements, where two UAV-BSs could meet users' requirements. By increasing the number of nodes with various service requirements, more UAV-BSs are needed as a UAV-BS cannot serve all of the nodes. Moreover, the coverage area of a UAV-BS is limited. 


Figs.~\ref{fig:100},~\ref{fig:200}, and~\ref{fig:500} illustrate the effects of the number of UAV-BSs, ground nodes, and the farm sizes on the CQI. Additionally, Table~\ref{tab:3scenario} provides information on the minimum number of UAVs required to serve all nodes' requirements in the farm. As farm sizes are various in nations, we considered three sizes: small, medium, and large. Small farms are quite common in China, whereas the size of farms is large in Australia \cite{lowder2016number}. The average number of herds based on the type of animals varies, e.g., sheep herds are much higher than beef and dairy herds \cite{beggs2019effects}. In various researches, the farms are clustered based on the number of heads in herds. However, we consider the farm size and the number of animals (nodes) that the UAV-BSs need to serve as per their requirements.

\begin{table}
\caption{Required number of UAVs by farm size and number of ground nodes to serve meeting their requirements. }
\centering
\scriptsize
\begin{tabular}{!{\color{black}\vrule}l|l!{\color{black}\vrule}l!{\color{black}\vrule}l|l|l|l!{\color{black}\vrule}} 
\hline
\multirow{2}{*}{Size of Farm~} & \multicolumn{6}{l!{\color{black}\vrule}}{Number of Ground Nodes}  \\ 
\cline{2-7}
                               & ~50 & 100~ & 200 & 500 & 800 & 1000~                        \\ 
\hline
~Small  & ~1  & ~1   & 2   & 2   & 4   & ~5    \\ 
\hline
~Medium                        & ~2  & ~3   & 6   & 8   & 12  & ~13                          \\ 
\hline
~Large                         & ~14 & ~18  & 20  & 21  & 24  & ~24                          \\
\hline
\end{tabular}
    \label{tab:3scenario}
\end{table}

\subsection{Meeting Nodes Requirements}

To evaluate our proposed method, we compared our results with those of the Chinese Restaurant Process (CRP), which is widely used for clustering~\cite{tsiropoulou2018interest}. In CRP, the clusters are formed based on three distinct factors, including IoT nodes' interest, physical proximity, and nodes' energy level. This paper adopts the CRP method for the scenario intending to evaluate and compare the methods' functionality. While the Ground nodes' service types are not the same, in the simulation, the types of services are not considered in nodes clustering. Subsequently, the CRP's interest coefficient would not affect nodes clustering. In CRP's method, nodes' distance coefficient profoundly affects the clustering. In addition, the concentration parameter, $a$, indicates the ground node's willingness to join other clusters or create a new cluster. Then, we assumed $a=2$, which has an impact on nodes clustering, and formed clusters have various ground nodes. 

As illustrated in Table~\ref{tab:comperinf}, we separate the farms into some clusters based on our proposed clustering method ($P_{Method}$) and the CRP method ($CRP_{Method}$). Next, a UAV is assigned to each cluster. Following this, the number of nodes served by a UAV is calculated with the aim of evaluating the functionality of the proposed method with $CRP_{Method}$. 

\begin{table}
\caption{Required number of UAV-BSs to serve nodes' meeting their requirements by the proposed MCDM method }
\centering
\small
\arrayrulecolor{black}
\begin{tabular}{!{\color{black}\vrule}l!{\color{black}\vrule}l!{\color{black}\vrule}l!{\color{black}\vrule}l!{\color{black}\vrule}} 
\hline
\multirow{2}{*}{Size of farms} & \multicolumn{3}{l!{\color{black}\vrule}}{Served Nodes / Number of UAVs}  \\ 
\cline{2-4}
                   & Small  & Medium  & Large    \\ 
\hline
                 & \textcolor{red}{98/1} & \textcolor{blue}{97/3} & 99/18 \\ 
$P_{Method}$     & 192/1  & \textcolor{orange}{196/5} & 193/20 \\ 
                 & 493/18  & 489/20 & 467/20 \\ 
\hline
                  & \textcolor{red}{99/11} & \textcolor{blue}{99/9} & 78/16
                  \\ 
$CRP_{Method}$    & 200/12 & \textcolor{orange}{174/11} & 154/15 \\
                  & 481/16 & 462/17 & 440/19 \\ \hline

\hline
\end{tabular}
    \label{tab:comperinf}
\arrayrulecolor{black}
\end{table}

Table~\ref{tab:comperinf} provides information concerning the number of served nodes by the UAV-BSs, in which their delay and bandwidth requirements are met. The number of nodes that are clustered by $CRP_{Method}$ is different, then the number of clusters is various. Subsequently, the UAV-BSs serve various numbers of nodes in clusters. Moreover, the number of UAV-BS assigned to the same scenario by the $CRP_{Method}$ is more than the $P_{Method}$. As an explanation, 100 ground nodes are clustered in 11 groups, including [29, 39, 7, 3, 4, 9, 3, 1, 1, 1, 3] nodes. Consequently, 11 UAV-BSs are assigned to the system, which is a lot more than 1 or 3 UAV-BSs assigned to the system by $P_{Method}$ in small and medium farms, respectively. It must be noted that in the largest farm, due to the sporadic activity and mobility of the animals, more UAV-BSs are required by the proposed clustering methods.

As we mentioned, the number of clusters in $CRP_{Method}$ is changed for various scenarios; for example, 100 nodes are grouped in 11 clusters in CRP, while these nodes are clustered to one cluster by $P_{Method}$ method. Despite using more number clusters in the $CRP_{Method}$ in comparison with the $P_{Method}$, the number of served nodes is approximately the same. Similarly, the performance of $P_{Method}$ in serving nodes by taking into account the number of utilized UAV-BSs is by far more considerable than the $CRP_{Method}$ in the medium farms by 100 ground nodes. It should be noted that the number of nodes in clusters by the $P_{Method}$ is distributed more balanced than $CRP_{Method}$ as nodes distribution in the $CRP_{Method}$ follows the concentration parameter. Subsequently, UAV-BSs meet more number of nodes by $P_{Method}$ in comparison to the $CRP_{Method}$ as illustrated for medium farms by 200 nodes. 

This paper clusters ground nodes by considering several parameters that significantly impact users' required services. As illustrated in Table~\ref{tab:comperinf}, our proposed method meets the user's requirements. Furthermore, a considerable time of this method is consumed for the clustering of the nodes by the k-means algorithm. The time complexity of this method is $O(n^2)$. 

\section{Conclusion}
\label{conclusion}




This paper proposed a multi-UAV service provisioning method to serve mobile users by the MCDM method. The MCDM method obtains an optimal number of UAVs in a smart farm with predefined dimensions, and animals tagged with IoT devices equipped with GPS sensors. The suggested method calculates an optimal number of UAV-BSs by considering the number of mobile users, required services, and the environment specifications for meeting users' requirements. In the MCDM method, the ground IoT nodes are classified into different clusters based on their locations and other attributes. Each IoT cluster is served by a group of UAV-BSs that follow strict rules to avoid collisions. Furthermore, the optimal number of UAVs decreases the service-providing cost, interference, and collision. Simulation-based performance evaluations demonstrated that the MCDM method outperforms the benchmark algorithm in meeting nodes' requirements, system efficiency, and reducing deployment costs.

\bibstyle{plain}
\bibliography{main}

\begin{thebibliography}{10}

\bibitem{SALEHI2021108415}
Shavbo Salehi and Behdis Eslamnour.
\newblock {Improving UAV Base Station Energy Efficiency for Industrial IoT
  URLLC Services by Irregular Repetition Slotted-ALOHA}.
\newblock {\em Computer Networks}, 199:108415, 2021.

\bibitem{our_ccnc_traj}
Sayed~Amir Hoseini, Ayub Bokani, Jahan Hassan, Shavbo Salehi, and Salil~S.
  Kanhere.
\newblock {Energy and Service-Priority aware Trajectory Design for UAV-BSs
  using Double Q-Learning}.
\newblock In {\em 2021 IEEE 18th Annual Consumer Communications Networking
  Conference (CCNC)}, pages 1--4, 2021.

\bibitem{salehi2019aetd}
Shavbo Salehi, Ayub Bokani, Jahan Hassan, and Salil~S Kanhere.
\newblock {Aetd: An Application-Aware, Energy-Efficient Trajectory Design for
  Flying Base Stations}.
\newblock In {\em 2019 IEEE 14th Malaysia International Conference on
  Communication (MICC)}, pages 19--24. IEEE, 2019.

\bibitem{OurISPN}
Shavbo Salehi, Jahan Hassan, Ayub Bokani, Sayed~Amir Hoseini, and Salil~S.
  Kanhere.
\newblock {Poster Abstract: A QoS-aware, Energy-efficient Trajectory
  Optimization for UAV Base Stations using Q-Learning}.
\newblock In {\em 2020 19th ACM/IEEE International Conference on Information
  Processing in Sensor Networks (IPSN)}, pages 329--330, 2020.

\bibitem{jahan_recharging}
Jahan Hassan, Ayub Bokani, and Salil~S Kanhere.
\newblock {Recharging of Flying Base Stations using Airborne RF Energy
  Sources}.
\newblock In {\em 2019 IEEE Wireless Communications and Networking Conference
  Workshop (WCNCW)}, pages 1--6. IEEE, 2019.

\bibitem{tran2020coarse}
Dinh-Hieu Tran, Thang~X Vu, Symeon Chatzinotas, Shahram ShahbazPanahi, and
  Bj{\"o}rn Ottersten.
\newblock {Coarse Trajectory Design for Energy Minimization in UAV-Enabled}.
\newblock {\em IEEE Transactions on Vehicular Technology}, 69(9):9483--9496,
  2020.

\bibitem{chang2020machine}
Zheng Chang, Wenlong Guo, Xijuan Guo, and Tapani Ristaniemi.
\newblock {Machine Learning-based Resource Allocation for Multi-UAV
  Communications System}.
\newblock In {\em 2020 IEEE International Conference on Communications
  Workshops (ICC Workshops)}, pages 1--6. IEEE, 2020.

\bibitem{zhao2020multi}
Nan Zhao, Zehua Liu, and Yiqiang Cheng.
\newblock {Multi-Agent Deep Reinforcement Learning for Trajectory Design and
  Power Allocation in Multi-UAV Networks}.
\newblock {\em IEEE Access}, 8:139670--139679, 2020.

\bibitem{zhang2020three}
Wenqi Zhang, Qiang Wang, Xiao Liu, Yuanwei Liu, and Yue Chen.
\newblock {Three-Dimension Trajectory Design for Multi-UAV Wireless Network
  with Deep Reinforcement Learning}.
\newblock {\em IEEE Transactions on Vehicular Technology}, 2020.

\bibitem{reina2018multi}
DG~Reina, Hissam Tawfik, and SL~Toral.
\newblock {Multi-subpopulation Evolutionary Algorithms for Coverage Deployment
  of UAV-networks}.
\newblock {\em Ad Hoc Networks}, 68:16--32, 2018.

\bibitem{8485481}
Lang Ruan, Jinlong Wang, Jin Chen, Yitao Xu, Yang Yang, Han Jiang, Yuli Zhang,
  and Yuhua Xu.
\newblock {Energy-efficient Multi-UAV Coverage Deployment in UAV Networks: A
  Game-theoretic Framework}.
\newblock {\em China Communications}, 15(10):194--209, 2018.

\bibitem{li2019reactive}
Xiaohui Li and Li~Xing.
\newblock {Reactive Deployment of Autonomous Drones for Livestock Monitoring
  Based on Density-based Clustering}.
\newblock In {\em 2019 IEEE International Conference on Robotics and
  Biomimetics (ROBIO)}, pages 2421--2426. IEEE, 2019.

\bibitem{zeng2019energy}
Yong Zeng, Jie Xu, and Rui Zhang.
\newblock {Energy Minimization for Wireless Communication with Rotary-wing
  UAV}.
\newblock {\em IEEE Transactions on Wireless Communications}, 18(4):2329--2345,
  2019.

\bibitem{lowder2016number}
Sarah~K Lowder, Jakob Skoet, and Terri Raney.
\newblock {The Number, Size, and Distribution of Farms, Smallholder Farms, and
  Family Farms Worldwide}.
\newblock {\em World Development}, 87:16--29, 2016.

\bibitem{beggs2019effects}
DS~Beggs, EC~Jongman, PH~Hemsworth, and AD~Fisher.
\newblock {The Effects of Herd Size on the Welfare of Dairy Cows in a
  Pasture-based system using animal-and resource-based indicators}.
\newblock {\em Journal of dairy science}, 102(4):3406--3420, 2019.

\bibitem{tsiropoulou2018interest}
Eirini~Eleni Tsiropoulou, Giorgos Mitsis, and Symeon Papavassiliou.
\newblock {Interest-aware Energy Collection \& Resource Management in Machine
  to Machine Communications}.
\newblock {\em Ad Hoc Networks}, 68:48--57, 2018.

\end{thebibliography}
\end{document}